\documentclass[a4paper,11pt]{article}
\pdfoutput=1 

\usepackage{jinstpub} 

\title{\boldmath Baby MIND: A magnetized segmented neutrino detector for the WAGASCI experiment}

\author[a,k]{M.~Antonova,}
\author[b]{R.~Asfandiyarov,} 
\author[c]{R.~Bayes,} 
\author[d]{P.~Benoit,}  
\author[b]{A.~Blondel,} 
\author[e]{M.~Bogomilov,} 
\author[f]{A.~Bross,} 
\author[b]{F.~Cadoux,} 
\author[g]{A.~Cervera,} 
\author[h]{N.~Chikuma,} 
\author[d]{A.~Dudarev,}  
\author[i]{T.~ Ekel\"of,}
\author[b]{Y.~Favre,} 
\author[a]{S.~Fedotov,}  
\author[c]{S-P.~Hallsj\"o,} 
\author[a]{A.~Izmaylov,} 
\author[b]{Y.~Karadzhov,} 
\author[a]{M.~Khabibullin,}
\author[a]{A.~Khotyantsev,} 
\author[a]{A.~Kleymenova,} 
\author[h]{T.~Koga,} 
\author[a]{A.~Kostin,} 
\author[a,j,k]{Y.~Kudenko,}   
\author[a]{V.~Likhacheva,} 
\author[b]{B.~Martinez,}
\author[e]{R.~Matev,}
\author[a]{M.~Medvedeva,}
\author[a,j,1]{A.~Mefodiev,\note{Corresponding author.}}
\author[l]{A.~Minamino,} 
\author[a]{O.~Mineev,} 
\author[d]{M.~Nessi,}
\author[b]{L.~Nicola,}
\author[b]{E.~Noah,}
\author[a]{T.~Ovsiannikova,} 
\author[d]{H.~Pais Da Silva,} 
\author[b]{S.~Parsa,}
\author[d]{M.~Rayner,}
\author[d]{G.~Rolando,}
\author[a]{A.~Shaykhiev,} 
\author[i]{P.~Simion,}
\author[c]{F.J.P.~Soler,}
\author[a]{S.~Suvorov,} 
\author[e]{R.~Tsenov,}
\author[d]{H.~Ten Kate,}
\author[e]{G.~Vankova-Kirilova}
\author[a]{and N.~Yershov.}

\affiliation[a]{Institute for Nuclear Research of the  Russian Academy of Sciences,\\60 October Revolution Pr 7a, Moscow, Russia}
\affiliation[b]{Dept. de Phys. Nucl. et Corpuscul. (DPNC), University of Geneva,\\Quai Ernest-Ansermet 24, Geneva, Switzerland}
\affiliation[c]{School of Physics and Astronomy, University of Glasgow,\\Kelvin Building, Glasgow, UK}
\affiliation[d]{European Organization for Nuclear Research, CERN,\\CH-1211 Geneva 23, Switzerland}
\affiliation[e]{Department of Physics, University of Sofia,\\James Bourchier Blvd. 5, Sofia, Bulgaria}
\affiliation[f]{Fermi National Accelerator Laboratory, \\Kirk Road and Pine Street Batavia IL 60510-5011, Illinois, USA}
\affiliation[g]{IFIC (CSIC $\&$ University of Valencia),\\Calle Catedr\`atico Jos\`e Beltran, 2, Valencia, Spain}
\affiliation[h]{International Center for Elementary Particle Physics, University of Tokyo,\\7-3-1 Hongo, Bunkyo-ku, Tokyo, Japan}
\affiliation[i]{University of Uppsala,\\752 36, Uppsala, Sweden}
\affiliation[j]{Moscow Institute of Physics and Technology,\\9 Institutskiy per., Dolgoprudny, Moscow Region, Russia}
\affiliation[k]{Moscow  Engineering Physics Institute,\\Kashirskoe shosse 31, Moscow,  Russia}
\affiliation[l]{Yokohama National University,\\79-8 Tokiwadai, Hodogaya, Yokohama, Japan}

\emailAdd{aleksandr.mefodev@cern.ch}

\abstract{ T2K (Tokai-to-Kamioka) is a long-baseline neutrino experiment in Japan designed to study various parameters of neutrino oscillations. A near detector complex (ND280) is located 280~m downstream of the production target   and  measures   neutrino beam parameters before any oscillations occur.  ND280's measurements are used to predict the number and spectra of  neutrinos  in the   Super-Kamiokande  detector at the distance of 295~km. The difference in the target material between the far (water) and near (scintillator, hydrocarbon) detectors leads to the main non-cancelling systematic uncertainty for the oscillation analysis. In order to reduce this uncertainty a new WAter-Grid-And-SCintillator detector (WAGASCI) has been developed. A magnetized iron neutrino detector (Baby MIND) will be used to measure momentum and charge identification of the outgoing muons from charged current interactions.   The Baby MIND modules are composed of magnetized iron plates and  long plastic scintillator  bars read out  at the both ends with wavelength shifting  fibers and silicon photomultipliers. The front-end electronics board has been developed  to perform the readout and digitization of the  signals from the scintillator bars. Detector elements  were tested  with cosmic rays and in the PS beam at CERN. The obtained results are presented in this paper.}


\collaboration[c]{}

\proceeding{International Conference "Instrumentation for Colliding Beam Physics" (INSTR17)\\
  27 February - 03 March, 2017 \\
  Novosibirsk, Russia}

\begin{document}
\maketitle
\flushbottom

\section{Introduction}
\label{sec:intro}
T2K (Tokai-to-Kamioka)~\cite{Abe:2011ks} is a long-baseline neutrino oscillation experiment in Japan which uses the Super-Kamiokande water Cherenkov detector as the far detector at the distance of  295 km from the neutrino beam source.  A near detector complex (ND280) is located  280~m away from the target and at $2.5^{\circ}$ with respect to the beam axis. ND280  consists of tracking detectors surrounded by an electromagnetic calorimeter, which are  placed inside a magnet. A large part of the ND280 detector is composed of plastic scintillator-based counters with wavelength shifting fibers.  The difference in the target material between  far (water) and near (scintillator, hydrocarbon) detectors is one of the major  non-cancelling systematic uncertainty in the T2K   oscillation analysis.

\begin{figure}[htbp]
\centering 
\includegraphics[width=.47\textwidth,origin=c]{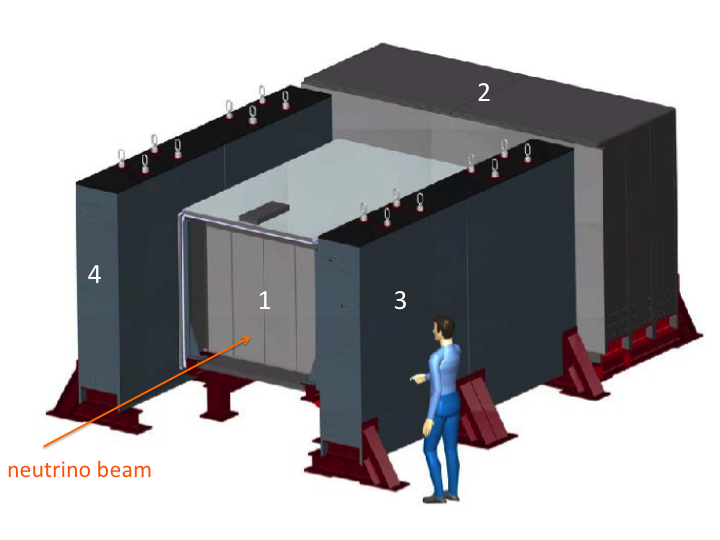}
\qquad
\includegraphics[width=.47\textwidth,origin=c]{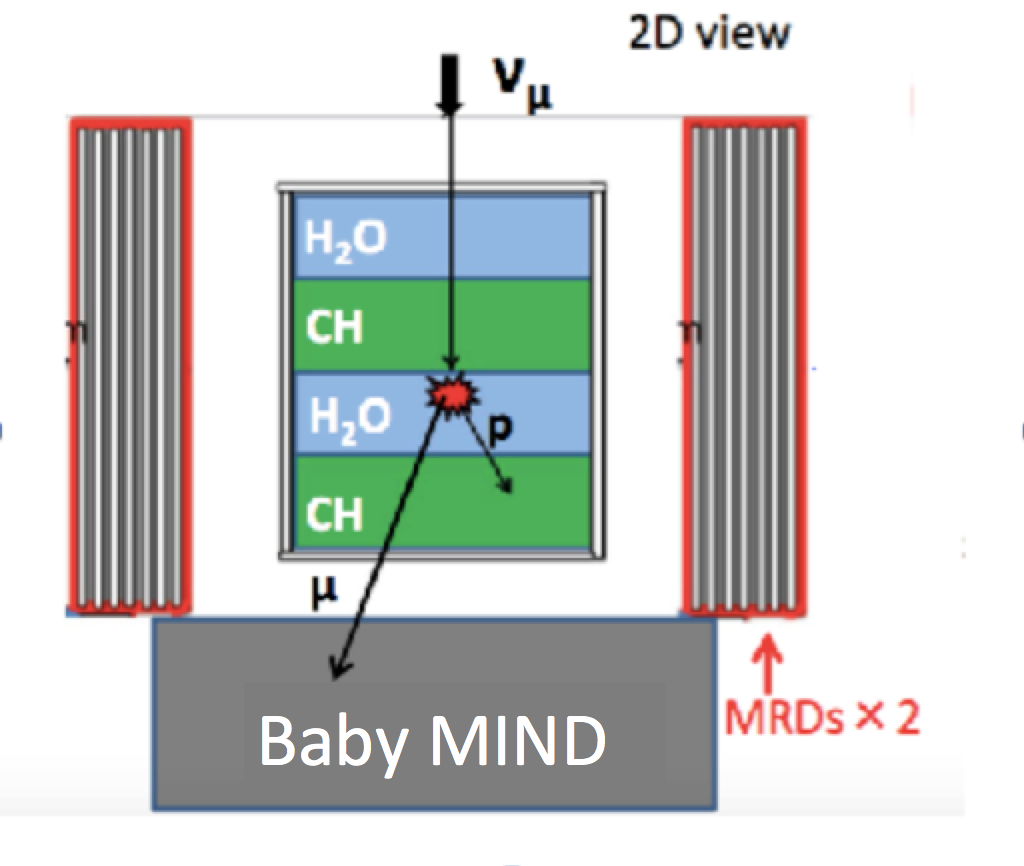}
\caption{\label{fig:WAGASCI}   Schematic view of the WAGASCI setup. {\it Left} --- {\bf 1}: 3D grid-structured WAGASCI water/scintillator neutrino target; {\bf 2}: Baby MIND; {\bf 3-4}: side MRD's. {\it Right} --- The event sample of neutrino interaction in the WAGASCI target is shown.}
\end{figure}

The WAGASCI experiment~\cite{Koga:2015iqa} at J-PARC has been proposed to reduce the systematic uncertainties  in  the neutrino cross-section measurements on water and hydrocarbon. The WAGASCI detector consists of two main parts, a 3D grid structured plastic scintillator target filled with $H_{2}O$ and muon range detectors (MRD's), which surround the central neutrino  target on two sides  (see figure~\ref{fig:WAGASCI}). 
The Magnetized Iron Neutrino Detector (Baby MIND)~\cite{BabyMIND, Antonova:2017tuf, {Antonova:2017thk}} is  located downstream of the neutrino target. The Baby MIND will provide charge identification and momentum measurements for muons resulting from neutrino interactions in the WAGASCI neutrino targets located
upstream. The charge of secondary particles from neutrino interactions are identified   by the curvature  of the particle track  in the Baby MIND magnetic field of 1.5 T. The MRDs   measure  the muon momentum by muon path length in the detector layers.  

\section{Baby MIND design}
\label{sec:Baby-MINDdesign}
\begin{figure}[htbp]
\centering
\includegraphics[width=1.0\textwidth,origin=c]{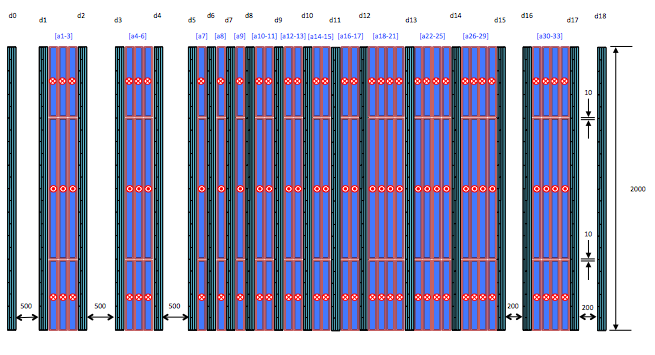}
\caption{\label{fig:BabyMINDdesign} The layout of Baby MIND. Scintillator planes are numerated from d0 (front plane) to d17, magnet plates are numerated  from s1 to s33. }
\end{figure}

\begin{figure}[htbp]
\centering
\includegraphics[width=.55\textwidth,origin=c]{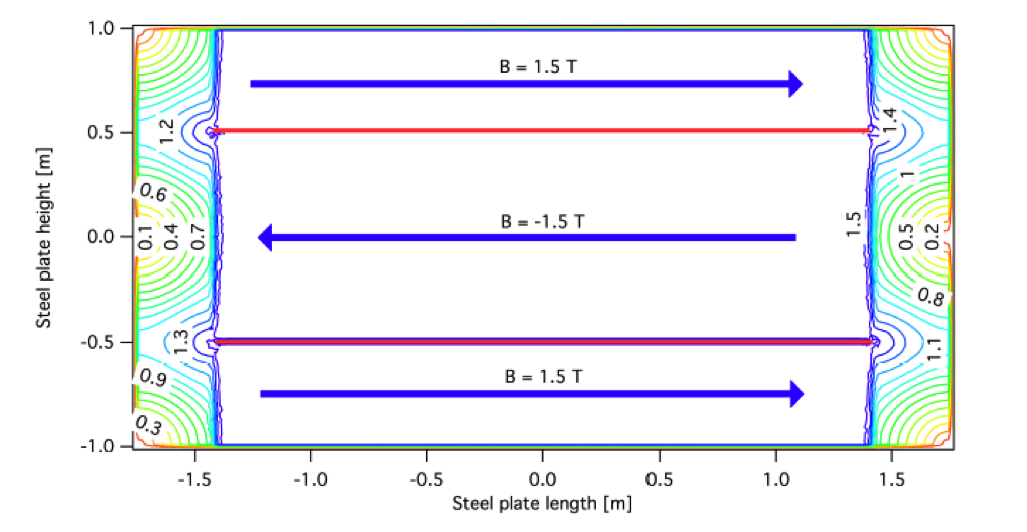}
\qquad
\includegraphics[width=.38\textwidth,origin=c]{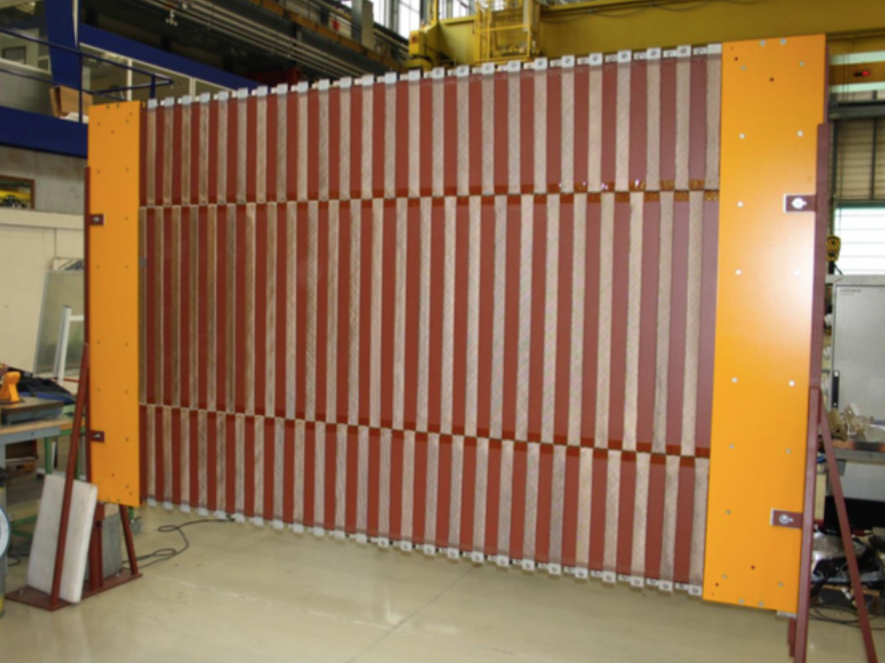}
\caption{\label{fig:Magnet} Two-slit configuration allows to halve the cross sections for flux return, leading to a larger central tracking region.}
\end{figure}

The main goal of the  Baby MIND is to identify     $\mu^+$/$\mu^-$ with a high  efficiency and to measure  their momenta in an interval of 0.3-5.0 GeV/c.    In this range the multiple scattering degrades the muon momentum measurements. Our simulations show that the  charge identification can be improved by optimizing the distance between the first magnet planes  for  the muon momentum down to 500 MeV/c.   Figure~\ref{fig:BabyMINDdesign} shows the basic design of the Baby MIND. The detector consists of  33 magnetized 30-mm thick iron plates and 18 scintillator planes. 
The magnet modules were designed to provide a minimum field of 1.5~T over the central tracking region with an uniformity of better than 10\% within the iron along the projection of a horizontal plastic scintillator bar. The stray field outside the iron plate is kept below 10~mT.
The total power consumption for 33 magnet modules is 12~kW~\cite{Magnet}. This is derived from the coil resistance and  the voltage drop across the power supply. To satisfy the space and transportation constraints, the design for the Baby MIND has individual magnetization of the modules by normal conducting coils wound on the surface of the iron plates. The magnet modules are made of the  ARMCO steel sheets with two horizontal slits machined in the center and are wrapped by an aluminum coil in a sewing pattern, as shown in Figure~\ref{fig:Magnet}.

\section{Scintillator modules}
\label{sec:Scintillatormodules}
The Baby MIND scintillator module (plane)  is assembled by gluing together  95 horizontal and 16 vertical scintillator bars which are read out with  WLS Kuraray Y11 fibers of 1~mm diameter and silicon photomultipliers  Hamamatsu  S12571-025C. The scintillator composition is a polystyrene doped with 1.5\% of paraterphenyl (PTP) and 0.01\% of POPOP. Vertical bar size is 0.7$\times$21$\times$195~cm$^3$, the WLS fiber covers the surface of scintillator in U-shape configuration with both fiber ends drawn at the same end of the bar. Horizontal bars are narrow and longer with 0.7$\times$3$\times$288~cm$^3$ size. The fiber is glued in the straight groove along the central axis so that readout is implemented at both ends of the bar.

In total, 1744 horizontal bars and 315 vertical bars were manufactured  at  the Uniplast company (Vladimir, Russia). The  0.7~cm thick extruded slabs  of 22~cm width were cut in the bars to the specified size, then the bars were covered by  30--100~$\mu$m thick diffuse reflector using etching of polystyrene surface with a chemical agent. Details about the manufacturing technology  can be found in~\cite{kudenko, mineev}.  


\section{Results of cosmic and beam tests}
\label{sec:Results}
Each bar was tested with cosmic rays to measure the  light yield at both ends~\cite{BMIND}. A trigger telescope of two large counters was installed  above and below of the central part of the  parallel array of 4 bars. The same Hamamatsu MPPCs S12571-025C  were utilized for the readout. The photosensor signal after preamplifier was sent to a digitizer CAEN DT5742 with 5~GHz sampling rate, then the charge was derived by integrating the signal pulse and calibrated in number of photoelectrons (p. e.). We did not apply the correction for MPPC cross-talk so the absolute light yield shown below is overestimated by 10\%. The light yield value was corrected for temperature variations during the tests with normalization  to  $T=22^\circ$C. 

\begin{figure}[htbp]
\centering
\includegraphics[width=.45\textwidth,origin=c]{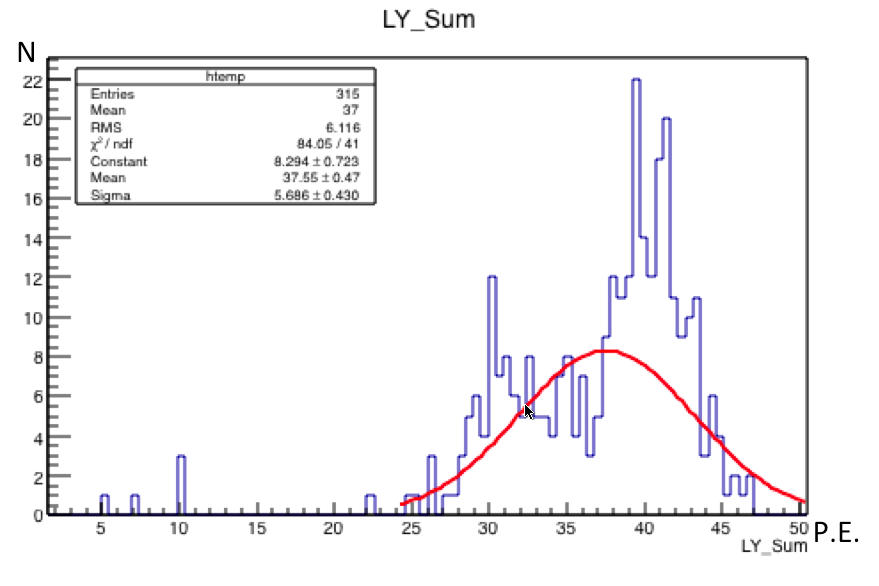}
\qquad
\includegraphics[width=.45\textwidth,origin=c]{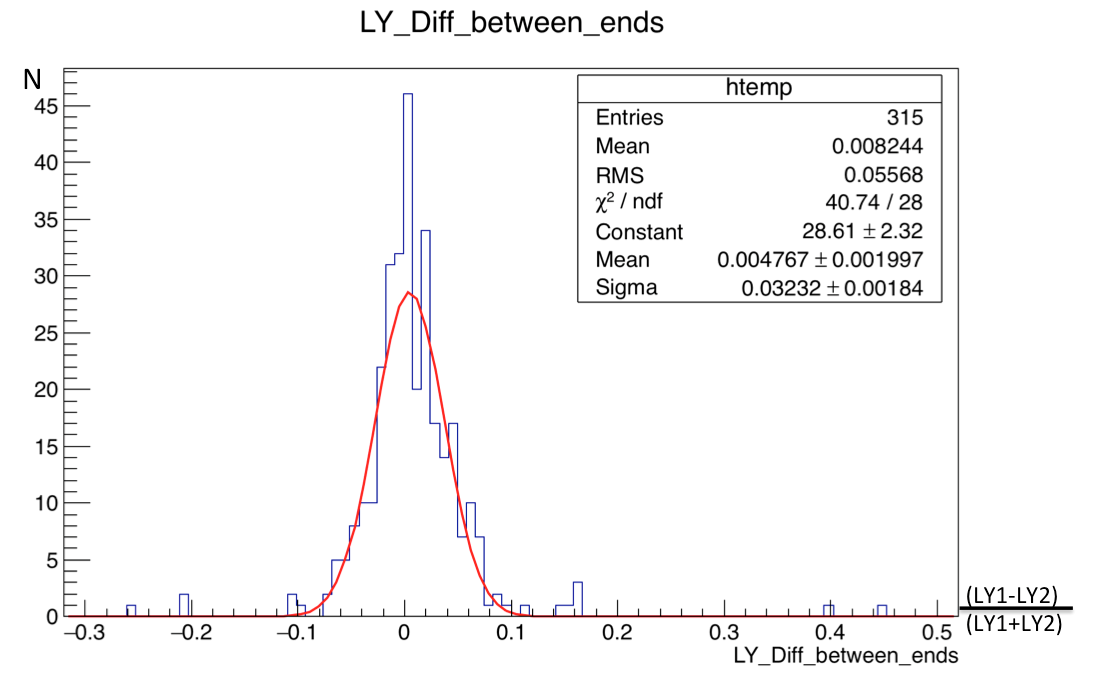}
\caption{\label{fig:vertical}  {\it Left} --- Distribution of the  light yield (sum of both ends) for 315 vertical bars, {\it right} --- distribution of the light yield asymmetry between the ends.}
\end{figure}

\begin{figure}[htbp]
\centering
\includegraphics[width=.45\textwidth,origin=c]{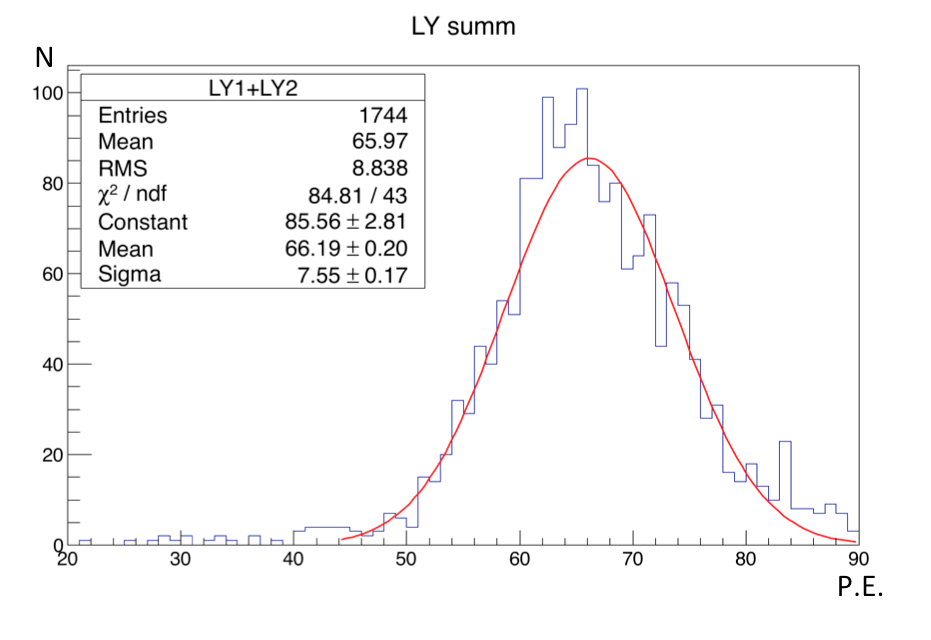}
\qquad
\includegraphics[width=.45\textwidth,origin=c]{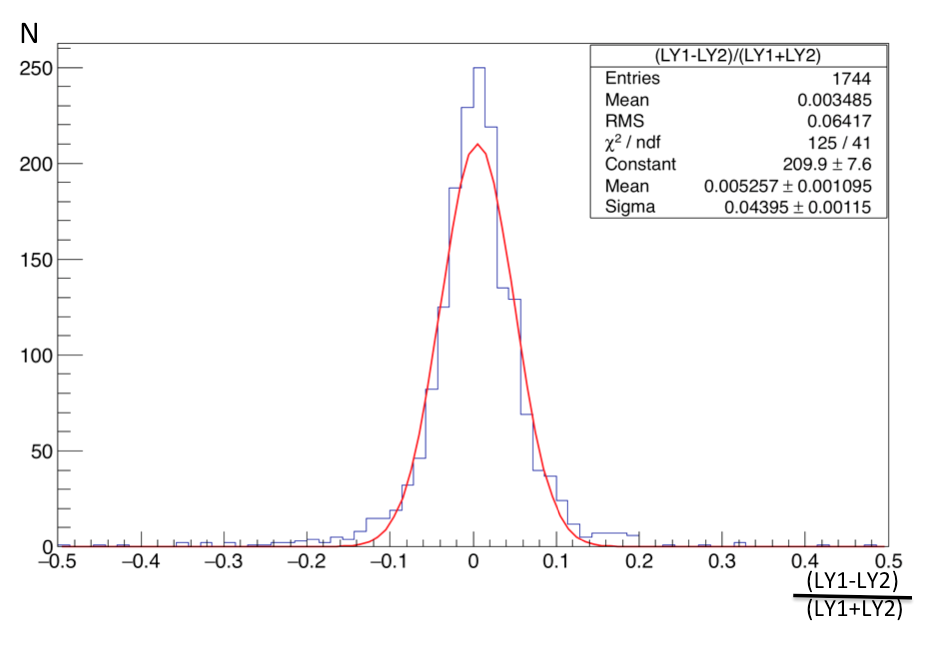}
\caption{\label{fig:horizontal} {\it Left} --- distribution of  light yield (sum from both ends) for 1744 horizontal bars, {\it right} --- distribution of the light yield asymmetry between the ends. }
\end{figure}

The average light yield (sum of both ends) was measured to be 37.5~p.e./MIP and 65~p.e./MIP for  vertical  and horizontal bars, respectively. Fig.~\ref{fig:vertical} 
demonstrates the distribution of the light yields for 315 vertical bars as well as the asymmetry in light yield between the ends.
Light yields and the asymmetry for 1744 horizontal bars are shown in Fig.~\ref{fig:horizontal}. 
Light yield asymmetry in both the horizontal and vertical bars  does not exceed 10\%.

\begin{figure}[htbp]
\centering
\includegraphics[width=.95\textwidth,origin=c]{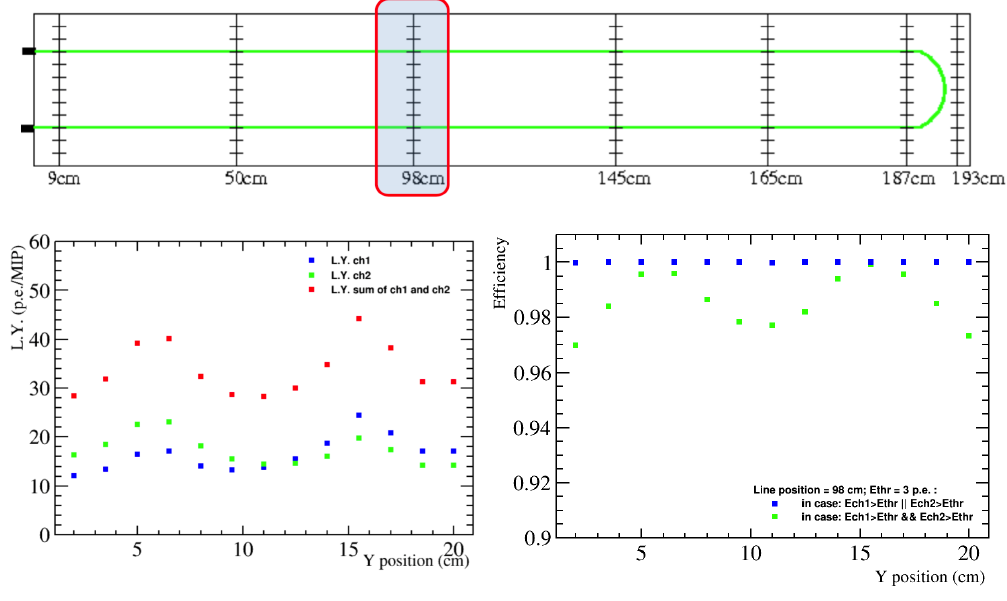}
\caption{\label{fig:effic} {\it Top} --- vertical  bar test map, {\it bottom left} --- light yield scan across a bar, {\it bottom right} --- detection efficiency scan across a bar (blue -- the events are counted if the charge is over the 3 p.e. threshold at either of the bar ends (OR logic), green --  if the charge is over the 3 p.e.  at each of the bar ends (AND logic).}
\end{figure}

Data obtained in the beam test at  the T9 beam area of  the CERN PS have been used to calculate the detection efficiency. The electronic trigger signal was produced by two trigger counters in coincidence. Figure~\ref{fig:effic} shows detection efficiency scan for vertical bars.  Efficiency and timing resolution for horizontal bars are described in details in a published paper~\cite{T9test}.

\section{Baby MIND software}
\label{sec:software}
Baby MIND uses the Simulation And Reconstruction Of Muons And Neutrinos (SaRoMaN) software suite \cite{Antonova:2017thk} to perform event reconstruction of both simulated muon and neutrino events as well as real beam data. The software suite was developed by the Baby MIND project and has been qualified with the Totally Active Scintillator Detector (TASD) \cite{BMIND} detector testbeam and will be further qualified with the Baby MIND testbeam. 

\section{Summary}
 The detector modules of Baby MIND have been assembled and prepared  for  beam tests  at CERN this year. The magnetization scheme and assembly procedure have been successfully proven through the construction  yielding 1.5 T in the magnet modules. The average light yield (sum from both ends) was measured with cosmic rays to be 37.5~p.e./MIP and 65~p.e./MIP for  vertical  and horizontal bars, respectively. 

A  new  Front  End  Board  based  on  the  CITIROC  ASIC  has  been  designed  for  the  readout of  scintillator  modules  that  will be used in  the  Baby MIND  muon  spectrometer.  Ongoing beam tests of Baby MIND modules with the front-end board  confirm a high light yield of the scintillator bars and  a successful operation of the electronic readout scheme.

\acknowledgments

The results of section \ref{fig:BabyMINDdesign} were obtained within European Union's Horizon 2020 Research and Innovation programme under grant agreement No. 654168. The results of section 4 were obtained within the RFBR/JSPS grant No. 17-52-50038 and the RFBR grant No.16-32-00766.



\begin{thebibliography}{99}

\bibitem{Abe:2011ks}
  K.~Abe {\it et al.} [T2K Collaboration],
  \emph{The T2K Experiment},
  \emph{Nucl.\ Instrum.\ Meth. A} {\bf 659} (2011) 106
  [arXiv:1106.1238 [physics.ins-det]].

 \bibitem{Koga:2015iqa}
  T.~Koga {\it et al.}
  \emph{Water/CH Neutrino Cross Section Measurement at J-PARC (WAGASCI 
   Experiment)},
  \emph{JPS Conf.\ Proc.}  {\bf 8} (2015) 023003. 

\bibitem{BabyMIND}
A.~Cervera {\it et al.} \emph{Performance of the MIND detector at a Neutrino Factory using realistic muon reconstruction}, \emph{Nucl.Instrum.Meth.} {\bf A624} (2010) 601-614.

\bibitem{Antonova:2017thk}
  M.~Antonova {\it et al.},
  \emph{Baby MIND: A magnetized spectrometer for the WAGASCI experiment,}
  arXiv:1704.08079 [physics.ins-det].

\bibitem{Antonova:2017tuf}
  M.~Antonova {\it et al.} [WAGASCI Collaboration],
  \emph{Baby MIND Experiment Construction Status,}
  arXiv:1704.08917 [physics.ins-det].

\bibitem{Magnet}
G.~Rolando {\it et al.} \emph{New and Optimized Magnetization Scheme for the Baby Magnetized Iron Neutrino Detector at J-PARC}, \emph{IEEE Transactions on Magnetics} {\bf 53} (2017) Issue 5.

\bibitem{kudenko}
Yu.~Kudenko {\it et al.} \emph{Extruded plastic counters with WLS fiber readout},  \emph{Nucl.Instrum.Meth.} {\bf 469} (2001) 340.

\bibitem{mineev}
O.~Mineev {\it et al.} \emph{Scintillator detectors with long WLS fibers and multi-pixel photodiodes}, \emph{Journal of Instrumentation} {\bf 6} (2011) P12004.
 
\bibitem{BMIND}
A.~Mefodiev {\it et al.} \emph{The design, construction and testing of Totally Active Scintillator Detector}, \emph{PoS} {\bf EPS-HEP2015:292} (2015) 067.
\bibitem{T9test}

W.~Baldini {\it et al.} \emph{Measurement of parameters of scintillating bars with wavelength-shifting fibres and silicon photomultiplier readout for the SHiP Muon Detector}, \emph{JINST} {\bf 12} (2017) P03005.




\end{thebibliography}
\end{document}